\documentclass[sigconf,nonacm]{acmart}

\usepackage{multirow} 
\usepackage{xcolor}
\usepackage{tabularx}
\usepackage{balance}

\newcommand{\pipeline}{{\texttt{GINGER}}}

\usepackage{tcolorbox}
\tcbuselibrary{listingsutf8}

\newtcolorbox{promptbox}{
  colback=gray!10,
  colframe=gray!50,
  boxrule=0.5pt,
  arc=2pt,
  left=4pt,
  right=4pt,
  top=4pt,
  bottom=4pt,
  fontupper=\ttfamily\small,
  width=\columnwidth
}

\AtBeginDocument{%
  }

\setcopyright{cc}
\copyrightyear{2025}

\begin{document}

\title{UiS-IAI@LiveRAG: Retrieval-Augmented Information Nugget-Based Generation of Responses}

\author{Weronika Łajewska}
\affiliation{%
  \institution{University of Stavanger}
  \city{Stavanger}
  \country{Norway}
}
\email{weronika.lajewska@uis.no}

\author{Ivica Kostric}
\affiliation{%
  \institution{University of Stavanger}
  \city{Stavanger}
  \country{Norway}
}
\email{ivica.kostric@uis.no}

\author{Gabriel Iturra-Bocaz}
\affiliation{%
  \institution{University of Stavanger}
  \city{Stavanger}
  \country{Norway}
}
\email{gabriel.e.iturrabocaz@uis.no}

\author{Mariam Arustashvili}
\affiliation{%
  \institution{University of Stavanger}
  \city{Stavanger}
  \country{Norway}
}
\email{mariam.arustashvili@uis.no}

\author{Krisztian Balog}
\affiliation{%
  \institution{University of Stavanger}
  \city{Stavanger}
  \country{Norway}
}
\email{krisztian.balog@uis.no}

\begin{abstract}
  Retrieval-augmented generation (RAG) faces challenges related to factual correctness, source attribution, and response completeness. The LiveRAG Challenge hosted at SIGIR'25 aims to advance RAG research using a fixed corpus and a shared, open-source LLM.
  We propose a modular pipeline that operates on information nuggets---minimal, atomic units of relevant information extracted from retrieved documents. This multistage pipeline encompasses query rewriting, passage retrieval and reranking, nugget detection and clustering, cluster ranking and summarization, and response fluency enhancement. This design inherently promotes grounding in specific facts, facilitates source attribution, and ensures maximum information inclusion within length constraints. 
  In this challenge, we extend our focus to also address the retrieval component of RAG, building upon our prior work on multi-faceted query rewriting. Furthermore, for augmented generation, we concentrate on improving context curation capabilities, maximizing the breadth of information covered in the response while ensuring pipeline efficiency. Our results show that combining original queries with a few sub-query rewrites boosts recall, while increasing the number of documents used for reranking and generation beyond a certain point reduces effectiveness, without improving response quality.
\end{abstract}



\maketitle

\section{Introduction}

The increasing reliance on conversational assistants such as ChatGPT for complex open-ended queries~\citep{Bolotova-Baranova:2023:ACL, Zamani:2023:FNT, Gabburo:2024:arXiv} presents challenges in factual correctness~\citep{Ji:2023:ACMa, Koopman:2023:EMNLP, Tang:2023:ACL}, source attribution~\citep{Rashkin:2021:Comput.}, information verifiability~\citep{Liu:2023:EMNLP}, consistency, and coverage~\citep{Gienapp:2024:SIGIR}. Although retrieval-augmented generation models aim to build responses based on retrieved sources~\citep{Lewis:2020:NIPS, Huang:2024:arXiv, Gienapp:2024:SIGIR}, they often struggle with transparency and source attribution. Current generative search engines frequently produce unsupported claims and inaccurate citations~\citep{Liu:2023:EMNLP}, underscoring the need for more reliable grounding. Although injecting evidence into prompts can mitigate hallucinations, long and redundant contexts can lead to the ``lost in the middle'' problem, where relevant information becomes inaccessible~\citep{Liu:2024:Trans.}. A post-retrieval refinement step is recommended to retain only essential details while preserving key information~\citep{Gao:2023:arXiv}.

\begin{figure*}[tp]
    \centering
    \vspace*{-0.25\baselineskip}
    \includegraphics[width=0.95\textwidth]{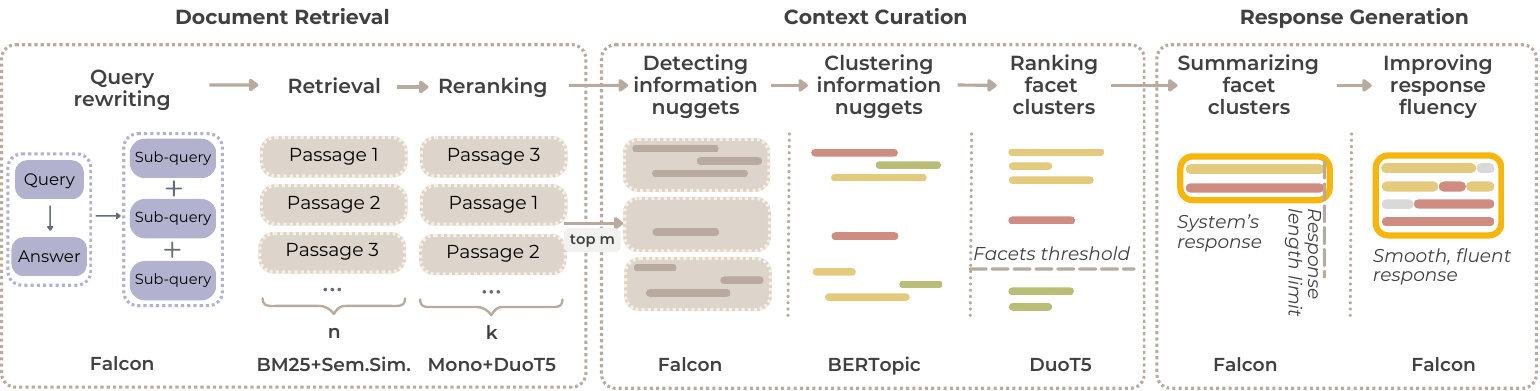}
    \captionsetup{aboveskip=5pt}
    \caption{High-level overview of our retrieval-augmented nugget-based response generation pipeline (\pipeline{}).}
    \label{fig:system_schema}
\end{figure*}

To address these limitations, we use a modular system for retrieval-augmented nugget-based response generation. It combines a strong retrieval pipeline with query rewriting, sparse and dense retrieval, and reranking with \textbf{G}rounded \textbf{I}nformation \textbf{N}ugget-Based \textbf{GE}nera\-tion of \textbf{R}esponses (\pipeline{})~\citep{Lajewska:2025:SIGIR} (see~Figure~\ref{fig:system_schema}). Unlike traditional RAG approaches, our method operates on atomic units of relevant information, called information nuggets~\citep{Pavlu:2012:WSDM}. Response generation involves identifying and clustering nuggets detected in retrieved passages, ranking clusters by relevance, summarizing them to eliminate redundancy, and then refining these summaries into a final, cohesive response. This process ensures comprehensive yet concise answers, maintains strong source attribution, and, as demonstrated in the TREC RAG'24 augmented generation task, significantly outperforms strong baselines. The core strength of GINGER lies in the granular, nugget-based processing of highly relevant information.

When developing our pipeline for the LiveRAG Challenge,\footnote{https://liverag.tii.ae/}, we conducted experiments on the TREC RAG’24 dataset as well as a small test dataset generated with DataMorgana~\citep{Filice:2025:arXiv}. Our results show that naive answer-based or single sub-question query rewriting can harm retrieval effectiveness, while combining the original query with a few diverse rewrites improves recall. Furthermore, optimizing reranking and generation parameters reveals that response quality improves only up to a point, beyond which sacrificing time efficiency yields limited gains. 

\section{Related Work}

Unlike traditional search engines that return a ranked list of documents, RAG systems provide a single, comprehensive response by synthesizing varied perspectives from multiple sources, blending the language fluency and world knowledge of generative models with retrieved evidence~\citep{Gienapp:2024:SIGIR, Mialon:2023:arXiv}. In retrieve-then-generate systems, generative processes are conditioned on retrieved material by adding evidence to the prompt~\citep{Izacard:2021:EACL, Shi:2023:arXiv, Ram:2023:Trans.} or attending to sources during inference. 

Systems submitted to the Retrieval-Augmented Generation track at the Text REtrieval Conference (TREC RAG'24)~\citep{Pradeep:2024:arXiv} have adopted modular architectures that improve the retrieval component by combining sparse and dense retrieval models, followed by reranking with models such as MonoT5 and DuoT5~\citep{Pradeep:2021:arXiv}, RankZephyr~\citep{Pradeep:2023:arXiv}, or other LLM-based graded relevance scoring. 
A notable enhancement involves query decomposition using an LLM to generate sub-questions, each addressing different facets of the information need. While LLM-based rewriting is well-established~\citep{Dhole:2024:ECIR, Weller:2024:ACL}, the generation of multiple diverse reformulations per query is a more recent development that shows strong potential for boosting recall and robustness by expanding the query's semantic coverage~\citep{Rackauckas:2024:IJNLC, Kostric:2024:SIGIR}. Retrieved and reranked results from these variants are typically merged using reciprocal rank fusion (RRF)~\citep{Wang:2024:TREC}.

For the generation stage, the most simplistic approach is to use proprietary models to generate responses in a single step based on the provided documents. However, ad hoc retrieval often returns documents with only partial relevance~\citep{Pavlu:2012:WSDM}, and placing relevant content in the middle of a long prompt can degrade generation quality~\citep{Liu:2024:Trans.}. While generative models often produce fluent and seemingly helpful responses, they frequently suffer from hallucinations and factual errors~\citep{Ladhak:2022:ACL, Ji:2023:ACMa, Liu:2023:EMNLP, Tang:2022:NAACL-HLT}. These limitations motivate more advanced context curation strategies, including unimportant token removal~\citep{Jiang:2023:arXiv}, content aggregation~\citep{Zhang:2024:arXiv}, and training extractors and condensers~\citep{Xu:2023:arXivb, Yang:2023:EMNLP}. 
Approaches at TREC RAG'24 include extracting, combining, and condensing the relevant information~\citep{Frobe:2024:TREC}, enhanced by verifying key facts across documents, rule-based redundancy removal, and enhancing coherence~\citep{Farzi:2024:TREC}. 

\section{Retrieval-Augmented Nugget-Based Response Generation}

Our approach, \pipeline{} (which stands for \textbf{G}rounded \textbf{I}nformation \textbf{N}ugget-Based \textbf{GE}neration of
\textbf{R}esponses), operates on information nuggets. It explicitly models various facets of the query based on retrieved information and generates a concise response that adheres to length constraints. It generates the response in three steps by: (1) retrieving top relevant passages from the corpus, (2) curating retrieved context for response generation, and (3) synthesizing the collected information into a final response; see Figure~\ref{fig:system_schema}. Our implementation adopts a modular architecture, with clearly separated components for each stage of the pipeline. This design allows for flexible experimentation and independent development of each component. All generation tasks, including query rewriting and context curation, are performed with the Falcon3-10B model\footnote{https://huggingface.co/tiiuae/Falcon3-10B-Instruct} accessed via the AI71 platform API.\footnote{https://ai71.ai/}

\subsection{Document Retrieval}

To reduce omissions caused by narrow queries, we apply query rewriting before retrieval. An LLM, queried without external documents, first generates a short answer to the original question. The assumption is that this intermediate answer surfaces the key aspects of the information need. We then ask the same model to generate $l$ additional queries, each focusing on a different aspect of that provisional answer while staying semantically consistent with the initial query.\footnote{Prompts used for query rewriting can be found in Appendix~\ref{app:prompts:query_rewriting}.} We combine each expanded query with the original, and then concatenate all $l$ rewrites together to create a final search string. Formally, 
$$
q' = (q + q_1') + \dots + (q + q_l')
$$
\noindent where $q$ is the original query, and each $q_i'$ is a rewrite focusing on a different aspect of the intermediate answer.

For retrieval, we adopt a two-stage retrieval pipeline, consisting of an initial passage retrieval step followed by re-ranking. First-pass retrieval is a combination of rankings obtained using both sparse and dense text representations. We use BM25 for sparse retrieval with an Opensearch-based index\footnote{https://opensearch.org/} and \texttt{intfloat/e5-base-v2}\footnote{https://huggingface.co/intfloat/e5-base-v2} embeddings with a Pinecone dense index.\footnote{https://www.pinecone.io/} Both indices are pre-built and provided by the challenge organizers. The retrieval results are then combined using reciprocal rank fusion~\citep{Cormack:2009:SIGIR}. For re-ranking, we first apply a pointwise re-ranker (\texttt{castorini/monot5-base\-msmarco}),\footnote{https://huggingface.co/castorini/monot5-base-msmarco} followed by a pairwise re-ranker (\texttt{castorini/duot5\-base-msmarco}),\footnote{https://huggingface.co/castorini/duot5-base-msmarco} both fine-tuned on the MS MARCO collection~\citep{Campos:2016:arXiv}, to refine the ranking and improve retrieval effectiveness.

\subsection{Context Curation}

Given the retrieved passages, \pipeline{} curates the context before the generation step to optimize response grounding and information relevance. First, we detect information nuggets within the top-$m$ ranked passages by prompting an LLM to annotate key information without altering the original text.\footnote{Prompts used for context curation can be found in Appendix~\ref{app:prompts:context_curation}.} Detected nuggets are then clustered according to different query facets to reduce redundancy and increase information density~\citep{Adams:2023:ACL}, leveraging the BERTopic model~\citep{Grootendorst:2022:arXiv}. Next, facet clusters are ranked for relevance using DuoT5 pairwise reranking ensuring that the most crucial clusters are prioritized for response generation~\citep{Gao:2023:arXiv, Liu:2024:Trans.}. This structured approach enables \pipeline{} to distill key information while preserving source attribution.

\subsection{Response Generation}

In the last step, \pipeline{} transforms the ranked facet clusters into a coherent response. Each top-ranked cluster is independently summarized into one sentence, following a prompt design that enforces conciseness and faithfulness to the original content~\citep{Goyal:2023:arXiv, Subbiah:2024:arXiv}.\footnote{Prompts used for response generation can be found in Appendix~\ref{app:prompts:res_gen}.} This modular summarization process ensures that the response remains factually accurate and grounded. However, since the response composed of independently summarized texts may lack fluency and coherence, we introduce a final refinement step where an LLM rephrases the response without introducing additional content. This ensures that the final output is not only factually reliable but also natural and readable, improving the overall user experience.

\subsection{Batch Processing Details}

To improve the efficiency of our pipeline, queries are processed in batches. 
We implemented multiprocessing with a concurrent queuing system, allowing each pipeline component to operate independently as long as its input queue is populated. This prevented bottlenecks and maximized hardware utilization. GPU-intensive components were distributed across 12 GPUs, with pointwise reranking and response generation using 25\% of total GPU resources and pairwise reranking the remaining 75\%. During the challenge day, we used 8 Tesla V100 GPUs and 4 NVIDIA A100 GPUs.

\section{Experiments}

In our experiments, we investigate our system's robustness with respect to the quality of the retrieved information. We also evaluate its ability to synthesize content from retrieved passages and reduce redundancy. The main goal of these experiments is to find a balance between efficiency---ensuring that responses can be generated for all test queries within a limited time window on the challenge day---and the quality of the generated responses.

\subsection{Datasets}

We generated a test set of 100 instances using the DataMorgana API, a synthetic benchmark generator platform used in the LiveRAG challenge~\citep{Filice:2025:arXiv}. DataMorgana enables RAG developers to create synthetic questions and answers from a given corpus based on configurable instructions. Half of the questions in our test set have answers grounded in a single document, while the other half are based on two documents. We experimented with several question categorizations proposed in the original paper, including factuality, premise, phrasing, and linguistic variation (see Table~\ref{tab:dm_dataset_categories} in Appendix~\ref{app:datasets}). Additionally, we incorporated the user expertise categorization and introduced two new categories for multi-document questions: comparisons between two entities and questions covering two aspects of the same topic. The documents provided by DataMorgana for each question are treated as ground-truth passages, and the generated answers serve as references to evaluate our system's responses.

We additionally employed the TREC RAG'24 dataset~\citep{Pradeep:2024:arXiv}, derived from the MS MARCO v2.1 collection and containing 301 information-seeking queries with graded relevance judgments. Unlike DataMorgana, which offers at most two judged passages per query, TREC RAG provides relevance labels for many candidate documents, giving a more reliable signal for retrieval evaluation. We used these judgments to benchmark the query rewriting component.

\subsection{Evaluation}

We evaluate the effectiveness of query rewriting primarily using the TREC RAG'24 dataset. The main metric is \emph{Recall@500}, computed using the trec\_eval tool.\footnote{https://github.com/usnistgov/trec\_eval} This cutoff corresponds to the number of top-ranked documents passed onto the pointwise reranker. We use the original query without any rewriting as the baseline.

For response generation, we use the AutoNuggetizer framework proposed for RAG evaluation and validated at TREC RAG'24~\citep{Pradeep:2024:arXiv}. AutoNuggetizer comprises two steps: nugget creation and nugget assignment. In nugget creation, nuggets are formulated based on relevant documents and classified as either ``vital'' or ``okay''~\citep{Voorhees:2004:NIST}. The second step, nugget assignment, involves assessing whether a system response contains specific nuggets from the answer key. 
The score \(V_{strict}\) for the system's response is defined as:
\[V_{strict}=\frac{\sum_{i}{ss_i^v}}{|n^v|} ~,\] 
where \(n^v\) represents the subset of the vital nuggets, and \(ss^v_i\) is 1 if the response supports the \emph{i}-th nugget and is 0 otherwise.
The score of a system is the mean of the scores across all queries.

\subsection{Results}

Results in Table~\ref{tab:rewriting_results} show that using a single rewrite alone underperforms even the original query, suggesting that naive rewriting can hurt retrieval effectiveness. While combining the original query with multiple rewrites improves recall, the gains saturate quickly. Adding more than three rewrites yields only marginal improvements, indicating diminishing returns beyond a small number of diverse reformulations. Notably, the recall achieved by using multiple rewrites alone is consistently lower than the recall obtained when those rewrites are concatenated with the original query, underscoring the importance of preserving the original formulation.\footnote{These experiments use TREC RAG data with a different retrieval collection, so the comparison to our pipeline is not direct. However, since we evaluate only the query rewriting component with retrieval frozen, the findings are expected to generalize to similar retrieval setups.} 

\begin{table}[tp]
    \centering
    \caption{Recall@500 for different query rewriting strategies on the TREC RAG'24 dataset. The best-performing configuration is shown in bold. \colorbox{teal!30}{Teal} background indicates the configuration used in the final submission.}
    \begin{tabular}{lc}
        \toprule
        \textbf{Rewriting Strategy} & \textbf{R@500} \\
        \midrule
         Original Query & 0.320 \\
         Single Rewrite & 0.217 \\
         Multi Rewrite (3) & 0.325 \\
         Multi Rewrite (10) & 0.357 \\
         Original Query + Single Rewrite & 0.343 \\
         \colorbox{teal!30}{Original Query + Multi Rewrite (3)} & 0.397 \\
         Original Query + Multi Rewrite (5) & \textbf{0.400} \\
         Original Query + Multi Rewrite (10) & 0.398 \\
         \bottomrule
    \end{tabular}
    \label{tab:rewriting_results}
\end{table}

Table~\ref{tab:eval_pipeline_variants} presents the evaluation of responses generated using different \pipeline{} configurations, assessed with the AutoNuggetizer framework. We varied two key parameters: the number of documents used for pairwise reranking ($k$) and the number of documents used for response generation ($m$). These parameters directly impact both the quality of the generated responses and the system’s efficiency. The reranking step with DuoT5 scales exponentially with $k$, while the number of Falcon API calls---dependent on $m$---is the main bottleneck in information nugget detection.

Given the two-hour time limit for processing 500 queries during the challenge (with three parallel processes), we aimed for a setup capable of handling at least 100 queries per hour. Although the setup with $k=50$ and $m=20$ produced the best responses, it exceeded our time constraints. Configurations with $k=40$, $m=10$ and $k=20$, $m=10$ yielded similar scores with much more efficient runtimes. Despite $k=20$ scoring slightly higher, we selected $k=40$ for our final submission to increase topic coverage and response diversity.

This choice is further supported by the limitations of AutoNuggetizer, which evaluates responses using nuggets extracted from only two documents. As a result, it may overlook relevant content captured by a broader reranking scope. In our manual analysis, we observed low scores for responses that were clearly grounded in relevant retrieved passages but where the available ground-truth nuggets were sparse. Conversely, high scores occurred mainly when our responses aligned exactly with the nuggets identified by AutoNuggetizer. This suggests that the framework’s effectiveness is constrained by its limited access to reference passages, which in turn restricts the evaluation of information quality.

\begin{table}[tp]
    \caption{Evaluation with AutoNuggetizer of responses generated with \pipeline{} using different setups. All variants use the top $n=500$ retrieved documents for pointwise reranking. \colorbox{teal!30}{Teal} background indicates the configuration used in the final submission.}
    \label{tab:eval_pipeline_variants}
    \centering
    \begin{tabular}{llcc}
        \toprule
        \textbf{Pairwise} & \textbf{Response} & \multirow{2}{*}{\textbf{V\_strict}} & \textbf{Time} \\
        \textbf{reranking} & \textbf{generation} & & \textbf{estimate} \\
        \midrule
        $k=50$ & $m=20$ & \textbf{0.406} & 70 min \\
        \colorbox{teal!30}{$k=40$} & \colorbox{teal!30}{$m=10$} & 0.397 & 41 min \\
        $k=20$ & $m=10$ & 0.404 & 42 min \\
        $k=20$ & $m=5$ & 0.350 & 26 min \\
        \bottomrule
    \end{tabular}
\end{table}

\subsection{Lessons Learned}

Participating in the LiveRAG challenge underscored the need to balance time efficiency with handling diverse query types. The time limit and the diversity of questions generated with DataMorgana posed unexpected challenges, requiring careful pipeline tuning and manual analysis.

Our initial query rewriting strategy, designed to sharpen the focus of the question using potential answer clues, worked well for factoid questions but underperformed for open-ended queries, where broader context is needed. This led us to revise our approach: using rewritten queries only for retrieval to ensure a diverse document pool, while letting reranking and generation rely on the original query to maintain relevance.

To meet the strict time window on challenge day, we had to rigorously optimize our system for efficiency. This involved extensive use of multiprocessing, batching, and distributing processes across multiple GPUs. The most resource-intensive component was the pairwise reranking stage, and the heavy reliance on the Falcon model across modules strained API rate limits. These constraints forced us to reduce the number of documents processed at each stage, carefully balancing efficiency against the quality of generated responses.

Finally, evaluating the responses with AutoNuggetizer surfaced key limitations of the framework. Its effectiveness depends on having a rich set of ground-truth nuggets derived from a broad set of relevant passages. In practice, especially for open-ended queries, this was often not the case, leading to unfairly low scores for responses that were, in fact, well grounded. This experience underlines the need for more robust response evaluation strategies, particularly when testing with limited access to ground-truth sources.
\section{Conclusions}

This paper has presented our participation in the LiveRAG Challenge at SIGIR'25, proposing a modular system for retrieval-augmen\-ted, nugget-based response generation. Our approach integrates query rewriting, sparse and dense retrieval, and reranking within the Grounded Information Nugget-Based Generation of Responses (\pipeline{}) framework. Evaluation on the TREC RAG'24 dataset and QA test samples from DataMorgana using the AutoNuggetizer framework demonstrates that our system effectively balances time efficiency and response quality.

\balance
\bibliographystyle{ACM-Reference-Format}
\bibliography{sigir2025-liverag}

\clearpage
\appendix
\section*{Appendix}

\section{Datasets}
\label{app:datasets}

Categorizations used in DataMorgana to generate our test samples are presented in Table~\ref{tab:dm_dataset_categories}.

\section{Prompts}

This section presents all the prompts used by our system for query rewriting, context curation and final response generation.

\subsection{Query Rewriting}
\label{app:prompts:query_rewriting}

Prompt for generating a concise answer to the query using the Falcon model:

\begin{promptbox}
\textbf{System:} You are a knowledgeable question answering AI that can answer a wide range of queries either in question form or keywords. \\
\textbf{User:} \{query\}.
\end{promptbox}

Prompt for rewriting query into a richer natural language variant:

\begin{promptbox}
\textbf{System:} You are a query rewriter that understands all necessary components of a good search query and helps users improve their queries. \\
\textbf{User:} Rewrite and return 3 query rewrites, each of which should cover a different aspect of the answer. The query rewrites should still be relevant to the original query. Return only the queries, one in each line. Do not add context, or any other information, or text.\\\\original query: \{query\}\\answer: \{answer\}
\end{promptbox}

\subsection{Context Curation}
\label{app:prompts:context_curation}

Prompt for detecting information nuggets in a passage given a query:

\begin{promptbox}
\textbf{System:} You are given a query and a relevant passage. Your task is to pinpoint and annotate the succinct excerpts within the passage that directly respond to the query. Ensure these excerpts are brief yet complete. Once identified, copy the entire passage and encapsulate the relevant snippets using <START> and </END> tags without changing any part of the original text. This includes avoiding modifications to words, punctuation, or formatting, as well as not adding any extra characters, symbols, or spaces. \\
\textbf{User:} Question: \{query\} Passage: \{passage\}
\end{promptbox}

\subsection{Response Generation}
\label{app:prompts:res_gen}

Prompt for summarizing an information cluster into a one-sentence-long text:

\begin{promptbox}
\textbf{System:} Summarize the provided information into one sentence (approximately 35 words). Generate one-sentence long summary that is short, concise and only contains the information provided. \\
\textbf{User:} \{information\_cluster\}.
\end{promptbox}

Prompt for improving the fluency of the generated response:

\begin{promptbox}
\textbf{System:} Rephrase the response given a query to improve its fluency. Do not change the information included in the response. Do not add information not mentioned in the original response. \\
\textbf{User:} Question: \{query\} Response: \{response\}
\end{promptbox}

\begin{table}
    \centering
    \caption{Categorizations used in DataMorgana to generate our test samples.}
    \footnotesize
    \renewcommand{\arraystretch}{1.2} 
    \setlength{\tabcolsep}{4.5pt}
    \begin{tabular}{p{1.05cm}p{1.25cm}p{5cm}} 
        \toprule
        \textbf{} & \textbf{Category} & \textbf{Description} \\
        \midrule
        \textbf{Factuality} & Factoid & Seeks a specific fact (e.g., date, number) \\
         & Open-ended & Invites elaborative or exploratory answers \\
        \midrule
        \textbf{Premise} & Direct & No premise or context about the user \\
         & With Premise & Includes short user-relevant background info \\
        \midrule
        \textbf{Phrasing} & Concise and Natural & Natural, direct questions (<10 words) \\
         & Verbose and Natural & Natural questions with more than 9 words \\
         & Short Search Query & Keyword-style, <7 words, no punctuation \\
         & Long Search Query & Keyword-style, >6 words, no punctuation \\
        \midrule
        \textbf{Linguistic Variation} & Similar to Document & Uses terms and phrasing from the source documents \\
         & Distant from Document & Uses different wording than the source documents \\
        \midrule
        \textbf{User Expertise} & Expert & Asks complex, domain-specific questions \\
         & Common Person & Asks basic, general-interest questions \\
        \midrule
        \textbf{Answer Type} & Multi-Aspect         & Covers two aspects of the same topic; needs info from two documents \\
         & Comparison    & Compares two entities; each described in separate documents \\
        \bottomrule
    \end{tabular}
    \label{tab:dm_dataset_categories}
\end{table}

\end{document}